# Spatial Topology and its Structural Analysis based on the Concept of Simplicial Complex

Bin Jiang[1] and Itzhak Omer[2]

[1]Department of Land Surveying and Geo-informatics, The Hong Kong Polytechnic University, Hung Hom, Kowloon, Hong Kong, Email: bin.jiang@polyu.edu.hk

[2]Department of Geography and Human Environment, The Environmental Simulation Laboratory, Tel-Aviv University, Ramat-Aviv, Tel-Aviv, 69978, Israel, Email: omery@post.tau.ac.il

**ABSTRACT**: This paper introduces a model that identifies spatial relationships for a structural analysis based on the concept of simplicial complex. The spatial relationships are identified through overlapping two map layers, namely a primary layer and a contextual layer. The identified spatial relationships are represented as a simplicial complex, in which simplices and vertices respectively represent two layers of objects. The model relies on the simplicial complex for structural representation and analysis. To quantify structural properties of individual primary objects (or equivalently simplices), and the simplicial complex as a whole, we define a set of centrality measures by considering multidimensional chains of connectivity, i.e. the number of contextual objects shared by a pair of primary objects. With the model, the interaction and relationships with a geographic system are modeled from both local and global perspectives. The structural properties and modeling capabilities are illustrated with a simple example and a case study applied to the structural analysis of an urban system.

## 1. INTRODUCTION

The first law of geography nicely describes an important nature of geographic systems in which "everything is related to everything else, but near things are more related than distant things" (Tobler 1970). The law suggests an overall structure of any geographic system, i.e. near things are more related than distant things. This is indeed true, as spatial phenomena are not random. More significantly, the law implies an underlying network view as to how "things" are interconnected. This network view is often represented as a graph in which the "things" are vertices and thing-thing relationships are edges. This can be seen clearly from the well known TIGER data model. Under the data model, linear objects form a kind of network in conventional quantitative geography (Haggett and Chorley 1969) and transportation modelling (Miller and Shaw 2001). Based on the geometric network representation, a series of network analyses can be carried out such as geocoding and routing analysis. A street network can also be perceived from a purely topological perspective (Jiang and Claramunt 2004), with which some universal patterns can be illustrated (e.g. Jiang 2007). In fact, the network analysis can be extended to all types of objects in Geographic Information Systems (GIS), as long as a rule of relationship is given for setting up an interconnected relationship or topology. Current GIS, mainly based on a geometric paradigm, have not yet adopted the topological paradigm for advanced spatial analysis (Albrecht 1997, Batty 2005), particularly in the context of the emerging science of networks (Barabási 2003, Watts 2004) following the two seminal papers (Watts and Strogatz 1998, Barabási and Albert 1999). Furthermore, most of the existing studies adopted a graph theoretic approach for setting up networks, and the thing-thing relationships are constrained by a direct relationship. For instance, two disjoint areas are usually considered to have no relationships.

This paper developed a model that can help to set up thing-thing relationships (for things which have no obvious direct relationships), in order to explore structural properties of individual objects of a geographic system, as well as the geographic system as a whole. Distinct from previous studies mostly based on graph theory for network representation, this paper adopted the concept of simplicial complex, as defined in the theory of Q-analysis (Atkin 1974), for the representation and structural analysis of geographic systems. The structural analysis is fundamental to understanding the flows of goods, information and people within a geographic system. Although Q-analysis shares with graph theory some similarity in terms of topology (Earl and Johnson 1981), the simplicial complex representation is capable of exploring structural properties from the perspective of multidimensional chains of connectivity. In this paper, we direct our attention to an overall structure or topology of an entire geographic system, and to how various parts (or objects) of the geographic system are interacted. In other words, our modeling effort is to take the interrelationship of spatial objects as a whole and explore its structure at both local and global levels.

Q-analysis, initially developed from algebraic topology and by Atkin (1974, 1977), is a set theoretic method to set up relationships of two sets and further explore the structure of the relationships. It provides a computational language for structural description using both geometric and algebraic representations. Because of its capability in structural description, Q-analysis has been used to explore man-environment interaction, ecological relation, and urban structure in social science, geography, and urban studies (Gould 1980). Recently, Q-analysis has also found some interesting applications in complexity study (Degtiarev 2000), database structure (Kevorchian 2003) and economic structure analysis (Guo, Hewings and Sonis 2003). To the best of our knowledge, Q-analysis has not been explored for spatial analysis in GIS, in particular in the way we suggested in the paper. We believe that the model we suggested in the paper can provide unique insights into the structure of various geographic systems through layered geographic information – a common data format of existing GIS.

The remainder of this paper is organized as follows. Section 2 introduces spatial topology that is a network representation of spatial relationships within a geographic system. Section 3 presents how Q-analysis in general and simplicial complex in particular can be used to characterize structural properties of a spatial topology. Section 4 suggests three centrality measures to characterize structural properties of individual objects within a geographic system. The introduced model and measures are applied to a case study for topological analysis of an urban system in section 5. Finally section 6 concludes the paper and points out possible future work.

## 2. SPATIAL TOPOLOGY: DEFINITION AND ILLUSTRATION

Here we introduce a model to identify spatial relationships of geographic objects via common objects. To introduce the model, let us start with an example from social networks. Two persons, A and B, are not acquainted in a general social sense as colleagues or friends, but they can be considered as "adjacent", as both, for instance, belong to the same geographic information community. The community or organization people belong to is a context that keeps people "adjacent". Extending the case into a geographic context, we can say that, for instance, two houses are adjacent if they are both situated along the same street, or within the same district. We can further state that a pair of objects sharing more common objects is more "adjacent". Now let us turn to the formal model for identifying such spatial relationships. The model aims to represent spatial relationships of various objects as a simplicial complex. To this end, some formal definitions are also presented in the context of GIS.

### 2.1 Defining Spatial Topology

We start with a definition of map layer according to set theory. Map layer is defined as a set of spatial objects at a certain scale in a database or on a map. For example, $M = \{o_1, o_2, ..., o_n\}$, or $M = \{o_i \mid i = 1, 2, ..., n\}$ denotes a map layer (using a capital letter) that consists of multiple objects (using small letters). The objects can be put into four categories: point, line, area, and volume objects in terms of basic graphic primitives. Note that the definition of objects must be appropriate with respect to the modeling purpose. For instance, a street layer can be considered a set of interconnected street segments, or an interconnected named street depending on the modeling purpose (Jiang and Claramunt 2004); a city layer can be represented as a point or area object depending on the representation scale.

A spatial topology ($T$) is defined as a subset of the Cartesian product of two map layers, L and M, denoted by $T = L \times M$. To set up a spatial topology, we need to examine the relationship of every pair of objects from one map layer to another. As distinct to topological relationships based on possible intersections of internal, external and boundary of spatially extended objects (Egenhofer 1991), we simply take a binary relationship. That is, if an object $\ell$ is within, or intersects, another object $m$, we say there is a relationship $\lambda = (\ell, m)$, otherwise no relationship, $\lambda = \varnothing$ [Note: the pair $(\ell, m)$ is ordered, and $(m, \ell)$ represents an inverse relation denoted by $\lambda^{-1}$]. The relationship can be simply expressed as "an object has a relationship to a contextual object". If a set of primary objects shares a common contextual object, we say the set of objects are adjacent or proximate. Thus two types of map layers can be distinguished: primary layer for the primary objects, with which a spatial topology is to be explored, and contextual layer, whose objects constitute a context for the primary objects. It is important to note that the contextual layer can be given in a rather abstract way with a set of features (instead of map objects). This way, the relationship from the primary to contextual objects can be expressed as "an object has certain features". For the sake of convenience and with notation $T = L \times M$, we refer to the first letter as the primary layer and the second the contextual layer.

The notion of spatial topology presents a network view as to how the primary objects become interconnected via the contextual objects. A spatial topology can be represented as a simplicial complex. Before examining the representation, we turn to the definition of simplicial complex (Atkin 1977). A simplicial complex is the collection of of relevant simplices. Let us assume the elements of a set A form simplices (or polyhedra, denoted by $\sigma^d$ where d is the dimension of the simplex); and the elements of a set *B* form vertices according to the binary relation λ, indicating that a pair of elements ($a_i$, $b_j$) from the two different sets *A* and *B*, $a_i \in A$ and $b_j \in B$, are related. The simplical complex can be denoted as $K_A(B;\lambda)$. In general, each individual simplex is expressed as a *q*-dimensional geometrical figure with *q+1* vertices. The collection of all the simplices forms the simplicial complex. For every relation λ it is feasible to consider the conjugate relation, $\lambda^{-1}$, by reversing the relations between two sets A and B by transposing the original incidence matrix. The conjugate structure is denoted as $K_B(A; \lambda^{-1})$.

A spatial topology can be represented as a simplicial complex where the simplices are primary objects, while vertices are contextual objects. Formally, the simplicial complex for the spatial topology $T = L \times M$ is denoted by $K_L(M;\lambda)$, where L represents the primary layer, and M the contextual layer, the relation between a primary object and contextual object $\lambda = (\ell, m)$. A spatial topology can be represented as an incidence matrix $\Lambda$, where the columns represent objects with primary layer and the rows represent the objects with contextual layer. Formally it is represented as follows,

$$\Lambda_{ij} = \begin{cases} 1 : if\ \lambda = (i,j) \\ 0 : if\ \lambda = \varnothing \end{cases} \quad [1]$$

The entry 1 of the matrix indicates that a pair of objects *(i, j)* respectively from the two different layers L and M (i. e. $i \in L$ and $j \in M$) is related, while the entry 0 represents no relationship between the pair of the objects. The incidence matrix $\Lambda_{ij}$ is not symmetric, as $\lambda \neq \lambda^{-1}$ in general.

Spatial topology is not intended to replace the existing topology or spatial relationship representation, but to extend and enhance the existing ones for more advanced spatial analysis and modelling. In this respect, the simplicial complex representation provides a powerful tool for exploring structural properties of spatial topology. Before introducing the structural analysis, we take a look at a simple example.

## 2.2 A Simple Example
Let us assume with environmental GIS, three pollution sources whose impact areas are identified, through a buffer operation, as a polygon layer. It is likely that the three polluted zones overlap each other. This constitutes a map layer, denoted by $Y = (y_j | j = 1,2,3)$. To assess the pollution impact on a set of locations with another map layer $X = (x_i | i = 1,2,...6)$, it is not sufficient to just examine which location is within which pollution zones. We should take a step further to put all locations within an interconnected context (a network view) using the concept of spatial topology. For instance, location $x_1$ is out of pollution zone of $y_2$, but it may get polluted through $x_2$ and $x_5$, assuming the kind of pollution is transmittable. Only under the network view are we able to investigate the pollution impact thoroughly.

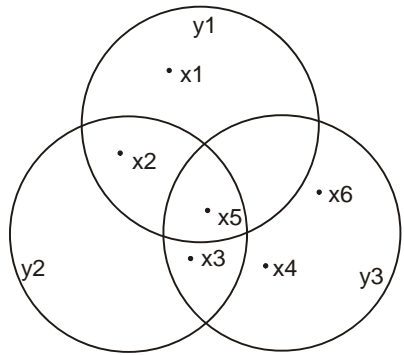

Figure 1: A simple example

In the example, the locations under pollution impact are of our primary interest. Using equation [1], the spatial topology can be represented as an incidence matrix as follows.

$$\Lambda_{36} = \begin{bmatrix} & x_1 & x_2 & x_3 & x_4 & x_5 & x_6 \\ y_1 & 1 & 1 & 0 & 0 & 1 & 0 \\ y_2 & 0 & 1 & 1 & 0 & 1 & 0 \\ y_3 & 0 & 0 & 1 & 1 & 1 & 1 \end{bmatrix}$$

For the primary layer, the six locations in figure 1 and with respect to the columns of the matrix can be represented as six simplices as follows:

$\sigma^0(x_1) = < y_1 >$
$\sigma^1(x_2) = < y_1, y_2 >$
$\sigma^1(x_3) = < y_2, y_3 >$
$\sigma^0(x_4) = < y_3 >$
$\sigma^2(x_5) = < y_1, y_2, y_3 >$
$\sigma^0(x_6) = < y_3 >$

where the right hand side of the equation represents vertices that consist of a given simplex. The dimension of a simplex is represented by a superscript. For instance, $\sigma^2(y_5)$ denotes a 2-dimensional simplex or a 2-dimensinal face that consists of three vertices $y_1$, $y_2$, and $y_3$. The simplices and the related complex can be graphically represented as figure 2.

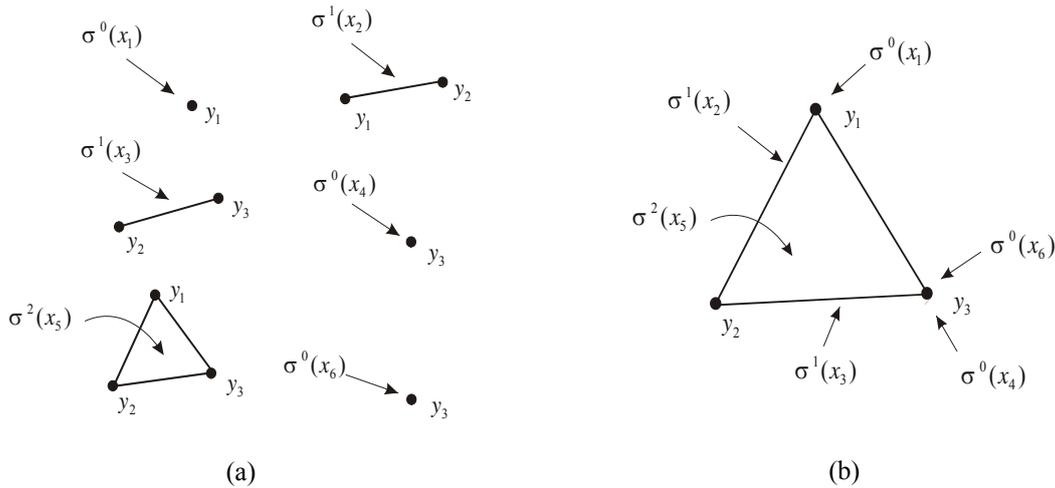

(a)　　　　　　　　　　　　　　　　(b)

Figure 2: The six simplices (a) and their collection – the simplicial complex Kx(Y; λ) (b)

Note that the primary and contextual objects are relative and transversal (interchangeable) depending on different application interests. If for instance we take Y as the primary layer (whether it makes sense is another issue which we will not consider here), i.e. to transpose the incidence matrix defined in equation [1], then we would have three simplices as follows (see also figure 3a),

$\sigma^2(y_1) = < x_1, x_2, x_5 >$
$\sigma^2(y_2) = < x_2, x_3, x_5 >$
$\sigma^3(y_3) = < x_3, x_4, x_5, x_6 >$

The collection of the three simplices constitutes a three-dimensional complex (figure 3b).

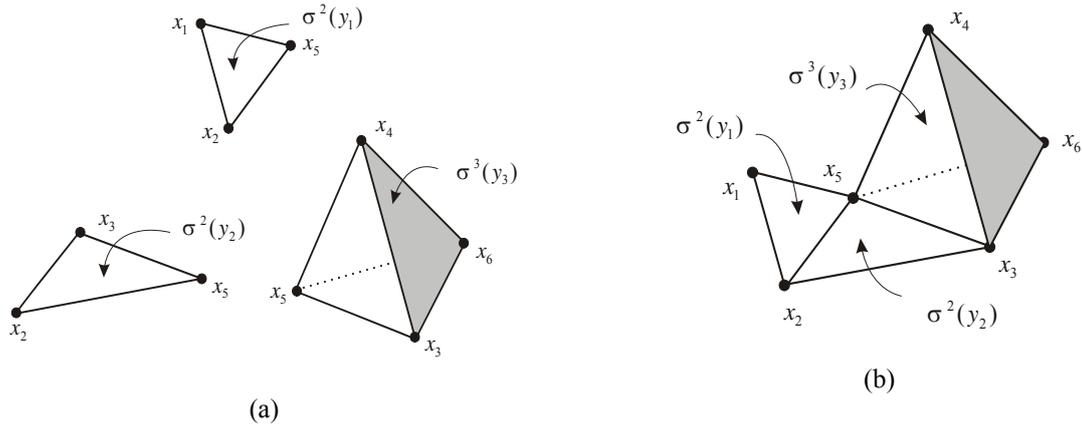

Figure 3: The three simplices (a) and their collection - the simplicial complex $Ky(X; \lambda^{-1})$ (b)

We have noted that a spatial topology can be represented as a geometric form (the simplicial complex). It is important to note that the geometric representation makes little sense when the dimension of the simplicial complex exceeds three because of a human perceptual constraint, but no so such constraint for the algebraic approach. In this paper, we mainly concentrate on the algebraic approach for structural analysis of a spatial topology.

## 3. STRUCTURAL PROPERTIES OF A SPATIAL TOPOLOGY

A spatial topology can be represented as a connected simplicial complex for structural analysis based on Q-analysis theory. In this case, primary objects are represented as individual simplices, with contextual objects as the vertices of simplices. For the primary layer, a pair of objects can be assessed to see if they share a common contextual object, and if so, how many common objects they share. If there are q+1 common contextual objects shared, we say the pair of primary objects is q-near. For example, 0-nearness means that a pair of primary objects shares one contextual object, and 1-nearness means 2 contextual objects are shared. For the example in figure 3 the pair $y_1$-$y_2$ shares two objects, i.e. two vertices $x_2$ and $x_5$, so we say $y_1$ and $y_2$ are 1-near. If, on the other hand, there is no contextual object shared, i.e. −1-near, but each of the primary objects is q-near one or several common primary objects, we say that the pair of objects is q-connected due to the transitive rule. For instance, the primary objects $y_1$ and $y_3$ have only one common contextual object ($x_5$), but they are 1-connected due to the primary object $y_2$ that serves as a transitional object; $y_2$ is 1-near with $y_1$ and with $y_3$. In a more general way, we say that two objects are q-connected if they are q-near or there is a chain of q-near simplices between them. That is, two objects can be connected through more than one transitional simplex (this is true of course for a case where there is more than 3 primary objects). As a general rule, two simplices are q-near, they would be also q-connected, but not vice versa.

The Q-analysis is based on the q-nearness and q-connectivity relations between the simplices of a given complex (or simplicial complex). A Q-analysis of a complex $K_L(M;\lambda)$ determines the number (#) of distinct equivalence classes, or q-connected components, for each level of dimension q ranging from 0 to q-1. The equivalence classes are determined by a rule as follows. If two simplices are q-connected (either q-near or q-connected), then they are in the same class. For example, the Q-analysis of the complex in figure 3 leads to the following equivalence classes at the different dimensional levels of q=0, q=1, q=2 and q=3. Each equivalence class is enclosed in the curly brackets.

q=0: $\{y_1, y_2, y_3\}$
q=1: $\{y_1, y_2, y_3\}$
q=2: $\{y_1\}, \{y_2\}, \{y_3\}$
q=3: $\{y_3\}$

The spatial topology can be analyzed by Q-analysis from a global perspective. That is, through counting the number of components at each q-level, we are able to get a structure vector Q. It is defined as follows:

$$Q = \{\#_{q-1}^{q-1} \ldots \#_2^2 \#_1^1 \#_0^0\} \qquad [2]$$

where the superscripts of the structural vector denote the dimension q, i.e. 0, 1, 2, …,(q-1) from right to left. Large entries in Q indicate that the structure at a certain dimension is fragmented or disconnected, while small entries imply more integrity structure. For instance, the structure vector for the complex in figure 3 $Q = \{\overset{3}{1} \; \overset{2}{3} \; \overset{1}{1} \; \overset{0}{1}\}$ indicates that there is one equivalence class (EC) at dimensional zero, one EC at dimension one, three ECs at dimension two, and one EC at dimension three. In other words, the complex is rather fragmented at the second dimension, and rather integrated at zero, first and third dimensions. Such an analysis can provide some critical insights into the structure of a spatial topology in terms of information flows.

From a local perspective we could assess the uniqueness of individual primary objects in the whole spatial topology by the eccentricity index. This index is defined by the relation between a dimension where an object is disconnected and another dimension where the object is integrated. It is formally expressed as follows.

$$Ecc(\sigma_i) = \frac{\hat{q} - \check{q}}{\check{q} + 1} \qquad [3]$$

where $\hat{q}$ or top-q denotes the dimension of the simplex ($\sigma_i$); $\check{q}$ or bottom-q denotes the lowest q level where the simplex gets connected to any other simplex. The eccentricity values of the primary objects in the complex presented in figure 3 are Ecc ($y_1$) = 0.5, Ecc ($y_2$) = 0.5, and Ecc ($y_3$) = 1.

According to equation [3], a primary object (as a simplex) gains a high value of eccentricity if it is different from every other. Thus it differentiates other simplices in the sense of uniqueness, rather than in the sense of importance or function in the flows and transmission within a complex. In other words, the eccentricity cannot precisely characterize the status of different simplices. This can be illustrated with the complex shown in figure 3, where both simplices $y_1$ and $y_2$ have the same eccentricity value (*0.5*). However, it is intuitive enough that $y_2$ tends to be more important structurally than $y_1$. Referring to the complex in figure 3, without $y_2$, the complex would be broken into two pieces at dimension two, while $y_2$ and $y_3$ are still kept together with the loss of $y_1$. The reason for this stems from the fact that the status of $y_2$, as a transitional simplex in the equivalence classes of dimensions of 0 and 1, is not taken into consideration in the eccentricity computation. In other words, the computation of eccentricity does not consider nearness relations within the equivalence class, but such relations are critical in determining the status of a simplex.

## 4. NEW MEASURES FOR CHARACTERIZING STRUCTURAL PROPERTIES OF INDIVIDUAL SIMPLICES

The inability of eccentricity to describe the status of individual simplices within a complex enables us to seek alternative measures for it. Centrality measures initially developed in the field of social networks (Freeman 1979) aim to support the quantitative description of a node status within a graph. It consists of three separate measures characterizing centrality from different perspectives: degree, closeness and betweenness. The degree centrality describes how many other nodes are connected directly to a particular node. It can be obtained by simply counting the number of directly linked nodes. The closeness quantifies how close a node is to every other node by computing the shortest distances from every node to every other. The betweenness measures to what extent a node is located in between the paths that connect pairs of nodes, and as such it reflects directly the intermediary location of a node along indirect relationships linking other nodes. We have extended the centrality measures into a simplicial complex to characterize the status of individual simplices within the complex. Basically, we define each measure at different dimensional levels, and then sum them up to show the centrality of individual simplices

### 4.1 Degree centrality

For describing the degree centrality of a given simplex within a complex, we must illustrate how many other simplices are directly connected to the simplex. We need to define the measure for a simplex at a different dimensional level. For a simplex ($\sigma_i$), the degree centrality at a given dimension level *q* can be formulated as follows:

$$D(\sigma_i)_q = \sum_{k=1}^{n} \lambda q(\sigma_i, \sigma_k) \qquad [4a]$$

where *n* is the total number of simplices within a complex, and $\lambda_q$ denotes whether there is direct relation ($\lambda$) between simplex ($\sigma_i$) to another simplex ($\sigma_k$) at dimension *q*.

To describe the overall degree centrality of a simplex within a simplicial complex, i.e., the total degree centrality of a simplex of different dimensions, we must accumulate the degrees at the different dimensions, i.e.,

$$D(\sigma_i) = \sum_{q=1}^{\max(q)} \sum_{k=1}^{n} \lambda q(\sigma_i, \sigma_k) \quad [4b]$$

where max(q) represents the highest level of dimension of the complex. Thus, the total degree centrality indicates to what extent a simplex is connected directly in the whole complex.

### 4.2 Closeness centrality

The closeness centrality in a graph context aims to assess the status of a node within a graph. It examines how a node is integrated or segregated with a graph or network. It is interesting that the measure has different names in different contexts. For instance, it is called status in graph theory (Buckley and Harary 1990); In space syntax (Hillier and Hanson 1984) it is called integration, computed from the reciprocal value of path length, a key concept used in small world theory (Watts and Strogatz 1998). Based on the idea of closeness centrality, we suggest here a similar measure to quantify how each simplex is close to every other. We define the measure for a simplex within a complex at the different dimensional level.

$$C(\sigma_i)_q = \frac{n-1}{\sum_{k=1}^{n} d(\sigma_i, \sigma_k)} \quad [5a]$$

where d is the shortest distance from a given simplex ($\sigma_i$) to every other simplex, n is the total number of simplices within a complex. Note the measure is given at a dimensional level q. Or put differently, the computation of the closeness considers multidimensional chains of connectivity. A simplex gains a closeness value at the different dimensional level if it gets connected to other simplices.

Logically the closeness measure can be accumulated along the different dimensional levels, to show an overall closeness of a simplex to every other at various dimensional levels. The accumulated closeness can be formulated as follows.

$$C(\sigma_i) = \sum_{q=0}^{\max(q)} \frac{n-1}{\sum_{k=1}^{n} d(\sigma_i, \sigma_k)} \quad [5b]$$

where max(q) represents the highest level of dimension of the complex.

### 4.3 Betweenness centrality

On the basis of the idea of betweenness centrality, we suggest here a similar measure for describing the linkage status of a simplex within a complex. Unlike the betweenness centrality in graph theory where the analysis focuses on zero-dimensional links using a term in q-analysis, we consider the multidimensional nature of the linkage. That is, to what extent a simplex is located in between the paths that connect pairs of simplices in the simplical complex at different dimensional levels and thereby determine the betweenness centrality of a given simplex at all dimensions. Accordingly, we define betweenness centrality for a simplex ($\sigma_i$) in a given dimension q as follows:

$$B(\sigma_i)_q = \sum_i \sum_j \frac{P_{ikj}}{P_{ij}} \quad [6a]$$

where $P_{ij}$ denotes the number of shortest paths between simplices $\sigma_i$ and $\sigma_j$, and $P_{ikj}$ is the number of shortest paths from $\sigma_i$ to $\sigma_j$ that pass through transitional simplex $\sigma_k$, i.e. the simplices $\sigma_i$ and $\sigma_j$ are q-connected by the transitional simplex $\sigma_k$; $P_{ikj}$ is a binary value that indicates if a simplex $\sigma_k$ serves as a transitional simplex for $\sigma_i$ and $\sigma_j$ - if so, $P_{ikj}=1$, otherwise $P_{ikj}=0$. Hence, the transitive weight of each transitional simplex $\sigma_k$ for a given pair $\sigma_i \sigma_j$ is equal to the proportion $1/P_{ij}$ i.e. if there is only one minimum path between i and j then $\sigma_k$ gains the value 1; if there are two minimum paths $\sigma_k$ gains the value ½.. Accordingly, the betweenness centrality of a simplex $\sigma_k$ at dimension level q is equal to the cumulative transitive weight values of $\sigma_k$ for all the pairs at that dimension. Note: the betweenness value does not exist when $P_{ij}=0$.

Applying the computation of $B(\sigma_i)_q$ for all the dimensions where the simplex is connected to other simplices defines the total betweenness centrality of the simplex within a complex K:

$$B(\sigma_i) = \sum_{q=0}^{\max(q)} \sum_i \sum_j \frac{P_{ikj}}{P_{ij}} \qquad [6b]$$

where max(q) represents the highest level of dimension of the complex. The total centrality measure is a cumulative value obtained by applying the computation of $B(\sigma_i)_q$ from bottom-q (the highest dimension at which a simplex gets connected to other simplices) down to dimension 0. As its counterpart defined in graph theory, this measure indicates to what extent a simplex is important for the linkage within a complex. If a simplex with higher value of betweenness gets removed, the whole complex may get broken into pieces.

### 5. STRUCTURAL ANALYSIS OF AN URBAN SYSTEM: CASE STUDY

We applied the introduced model and measures into the city of Tel Aviv in Israel for studying park-park and neighborhood-neighborhood relationships - a kind of structural analysis of urban systems. To achieve the goal, we first adopt neighborhoods as the contextual layer to assess the park-park relationships. That is, two parks are inter-related if the two are accessible by some neighborhoods. Depending on the number of neighborhoods, the strength of relationship varies from one pair to another, thus providing a kind of multiple chain of connectivity among the parks. In the case study, a layer of 46 parks (0-45), denoted by P, and a layer of 62 neighborhoods (0-61), denoted by N, are overlapped (Figure 4), create the simplical complex $K_p(N;\lambda)$ to assess whether a park is accessible by a neighborhood. It should be noted that parks are defined as the green areas bigger than 20,000 square meters, and a park is accessible by a neighborhood if it is within 0.5 km radius from the border of the park (these kinds of criteria are used for assessment of accessibility to the urban parks in Israeli cities). Based on the incidence matrix, we could then carry out a series of structural analysis for characterizing the individual parks with the introduced model and measures. With the case study, we will demonstrate some park-park relationships or patterns that are uniquely illustrated by our analysis. However the conjugate relation, of neighborhood-neighborhood relationship, $K_N(P; \lambda^{-1})$, have also potential to help identify a groups of connected neighborhoods via the urban parks. Namely, the urban parks enable direct and indirect relation or transmission between residents who belong to different neighborhoods.

Based on the above accessibility assessment, we generate an incidence matrix showing neighborhood-park relationship. According to equation [1], if park $P_i$ is accessible by neighborhood $N_j$, then the corresponding entry of the matrix is 1, otherwise is 0. Figure 5 illustrates the incidence matrix, whose 1s and 0s are indicated by black and white cells. Each park $P_i$, corresponding to the column of the matrix, can be considered as a simplex in $K_p(N;\lambda)$, thus the 46 parks form 46 simplices. The following is a partial list of the simplices:

$\sigma^6(P_0) = <N_{35}, N_{37}, N_{38}, N_{39}, N_{41}, N_{42}, N_{43}>$
$\sigma^4(P_1) = <N_{43}, N_{47} >$
…
$\sigma^5(P_{44}) = <N_1, N_2, N_3, N_4, N_6, N_7>$
$\sigma^0(P_{45}) = <N_1 >$

Obviously some simplices can be represented geometrically as a tetrahedron, while most others have the dimension of more than three. The higher dimensional simplices if represented geometrically are difficult to perceive by human beings in terms of the complexity of structure, particularly for a large system. In this case, the algebraic method of spatial topology shows an advantage in discovering structure from both local and global perspectives.

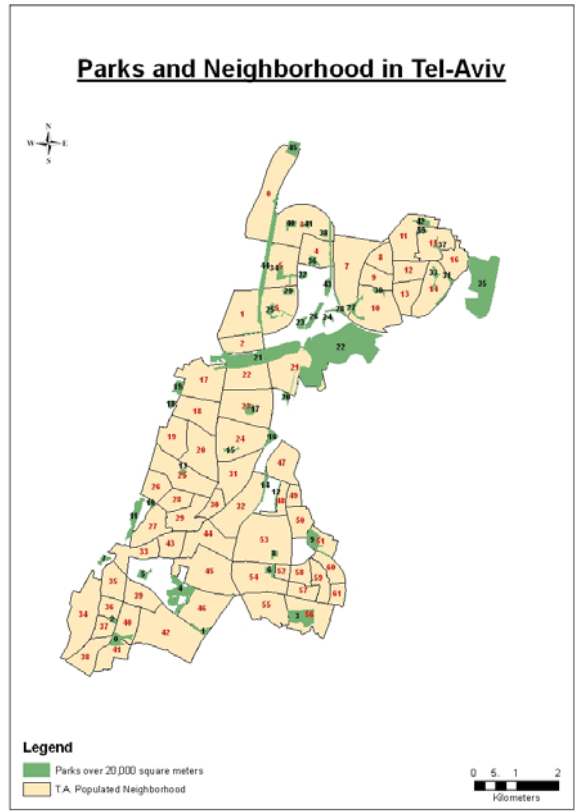

Figure 4: Parks and neighborhoods in Tel-Aviv

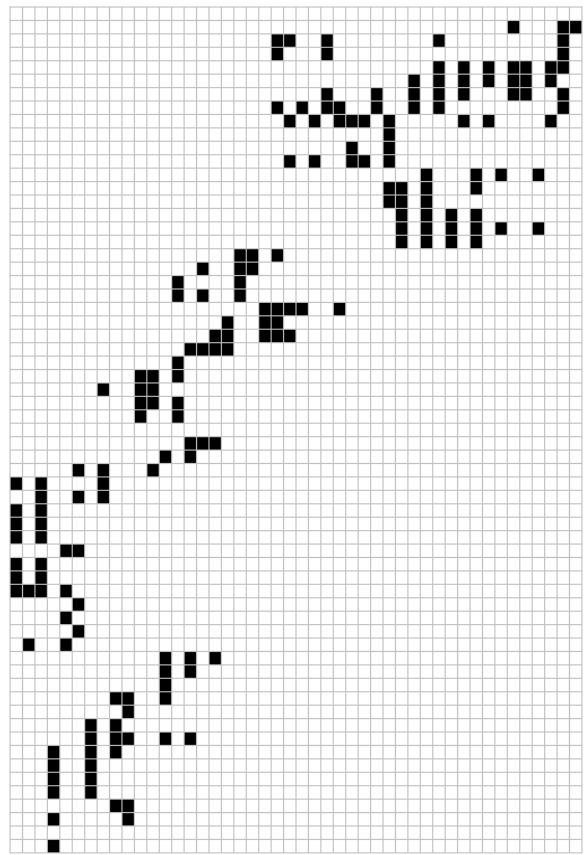

Figure 5: The incidence matrix represented by a raster format (black cell = 1, white cell = 0)

We then rely on the incidence matrix for Q-analysis, forming the connected components (or EC as previously mentioned) at different dimensions (see Table 1, where each component is enclosed by curly brackets). It is intriguing to note from the table that all the simplices at dimension zero (q=0) are interconnected, forming one interconnected component. In a contrast, the simplices are fragmented into two isolated parts at dimension one (q=1). Furthermore at dimension one, there are two interconnected components: the first containing 43 simplices, and second containing 2 simplices. In terms of Q-analysis, simplices within each component are either q-near or q-connected. For the biggest component at dimension one, it means that all the simplices within the component are 1-connected. The resulting connected components have far reaching implications in terms of structural analysis. From the structural perspective, the simplices at dimension one are not interconnected as a whole as at dimension zero. However, the interconnected components at dimension one tend to trade more flows among the individuals. Taking the pair of parks {1, 4} for example, it has two neighborhoods accessible to the pair. In a similar fashion, the two pairs {0, 2}, {27, 30} at dimension two (q=2) have stronger interaction since they have three neighborhoods accessible to each other. This kind of information is certainly unique to the multidimensional topological analysis suggested here. As such, this approach can be considered as an advantage in comparison to the traditional network analysis which did not consider the multidimensional chain of connectivity.

Table 1: Q-analysis of the park-park relationship
(NOTE: connected components are highlighted)

q = 0:   **{0,1,2,3,4,5,6,7,8,9,10,11,12,13,14,15,16,17,18,19,20,21,22,23,24,25,26,27,28,29,30,31,32,33,34,35,36,37,38,39,40,41,42,43,44,45}**

q = 1:   **{0,2,3,5,6,7,8,9,10,11,12,13,14,15,16,17,18,19,20,21,22,23,24,25,26,27,28,29,30,31,32,33,34,35,36,37,38,39,40,41,42,43,44}** **{1,4}**

q = 2:   **{0,2}** **{3,6,8,9}** {4} {5} {7} **{10,11,13}** **{12,14,16}** {15} {17} {18} **{20,21,25,32,34,36,38,40,41,43,44}** {22} {26} **{27,30}** **{31,33,35,37}**

q = 3:   **{0,2}** **{3,6}** {4} {5} {7} {8} {9} {10} {11} **{12,14}** {13} {15} {16} {18} {21} {22} **{25,34,44}** {30} **{31,33,37}** {36} {40}

q = 4:   **{0,2}** {3} {5} {6} {8} {9} {12} {13} {14} {21} {30} **{31,33,37}** {34} {44}

q = 5:   **{0,2}** {3} {6} {12} {13} {14} {21} {30} {33} {44}

q = 6:   **{0,2}** {21}

q = 7:   {2}

$Q = \{\overset{7}{1}\ \overset{6}{2}\ \overset{5}{10}\ \overset{4}{14}\ \overset{3}{21}\ \overset{2}{15}\ \overset{1}{2}\ \overset{0}{1}\}$

The structural vector $Q = \{\overset{7}{1}\ \overset{6}{2}\ \overset{5}{10}\ \overset{4}{14}\ \overset{3}{21}\ \overset{2}{15}\ \overset{1}{2}\ \overset{0}{1}\}$ according to equation [2] indicates the degree of integration or fragmentation at different dimensions for the individual parks. The results of the above analysis are mainly carried out at the global level.

From the local perspective, we can compute a series of measures including eccentricity, and the newly introduced centrality measures for individual simplices, to examine their status in their connected wholes. According to the definition of eccentricity, we can remark that a simplex (equivalent to a park) with eccentricity of zero is fully embedded in, or fully overlap with other simplices in the simplicial complex. On the contrary, a simplex with a higher eccentricity (non-zero) value tends to be distinctive or eccentric in the complex. Figure 6(d) illustrates the pattern of eccentricity of the parks. If a park has a very high connectivity, and on the other hand the park has a low number of accessible neighborhoods with any other park, then the park tends to by very distinctive or eccentric. From an operative perspective, it is important to keep the eccentric parks since they are critical for a provision of this service for certain urban neighborhoods. As mentioned early in the paper, the new centrality measures surpass the eccentricity in characterizing status of individual simplices of a complex. The introduced centrality measures consider multidimensional chains of connectivity at the different dimensions, so they comprehensively reflect the status of individual simplices within their complex. Compared to their counterparts in graph theory, the simplicial complex representation provides some unique insights into structural analysis. Figure 6 (a, b and c) illustrates the centrality measures for the individual parks.

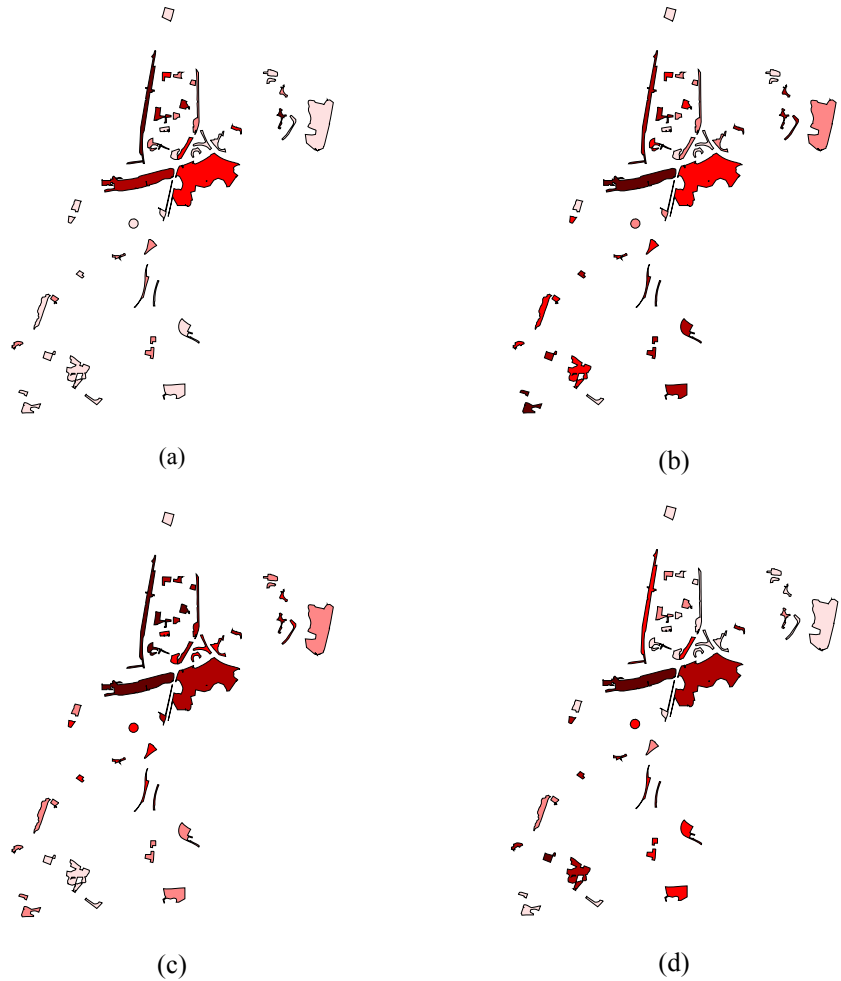

Figure 6: Betweenness (a), degree (b), closeness (c) and eccentricity (d) of the parks
(Note: values increase from light red to dark red)

The above analysis can be applied to the case of neighborhood-neighborhood relationship (Table 2), and insights towards the structure of the relationship can be obtained. In this case, neighborhoods constitute the primary layer, while parks are the contextual layer. Various structural patterns are illustrated in Figure 7. This analysis can help us to identify neighborhoods which have exclusive set of parks (high eccentricity), neighborhoods which are well connected (high degree), neighborhoods which have important role for information flow (high betweenness), and neighborhoods which are closer to other neighborhoods (high closeness).

Table 2: Q-analysis of the neighborhood-neighborhood relationship
(NOTE: connected components are highlighted)

| | |
|---|---|
| q = 0: | **{0,1,2,3,4,5,6,7,8,9,10,11,12,13,14,15,16,17,18,19,20,21,22,23,24,25,26,27,28,29,31,32,33,34,35,36,37,38,39,40,41,42,43,44,45,46,47,48,49,50,51,52,53,54,55,56,57,58,59,61}** |
| q = 1: | **{0,1,2,3,4,5,6,7,9,10,21,22,23,24,31,32,47,48,50,52,53,54,55,56,57,58}** **{11,12,13,14,15,16}** **{17,18,19,20}** **{26,27,28,29,33,34,35,36,37,38,40,41,42,46}** {39} {59} |
| q = 2: | {0} **{1,2,3,4,5,6,7,10,21,22,23}** **{11,12,13,14,15,16}** {17} {18} {20} **{24,31}** **{26,28}** {27} {33} {34} {35} {42} |
| q = 3: | **{1,3,4,5,6} {7,10} {11,14,15,16}** {12} {21} {23} {24} {42} |
| q = 4: | **{3,4,5,6} {7,10}** {15} {21} {23} {53} |
| q = 5: | **{3,4}** {5} {6} {7} {15} |
| q = 6: | {3} {4} {5} {6} {7} |
| q = 7: | {5} {6} {7} |
| q = 8: | {7} |

$$Q = \{\overset{8}{1}\ \overset{7}{3}\ \overset{6}{5}\ \overset{5}{5}\ \overset{4}{6}\ \overset{3}{8}\ \overset{2}{13}\ \overset{1}{6}\ \overset{0}{1}\}$$

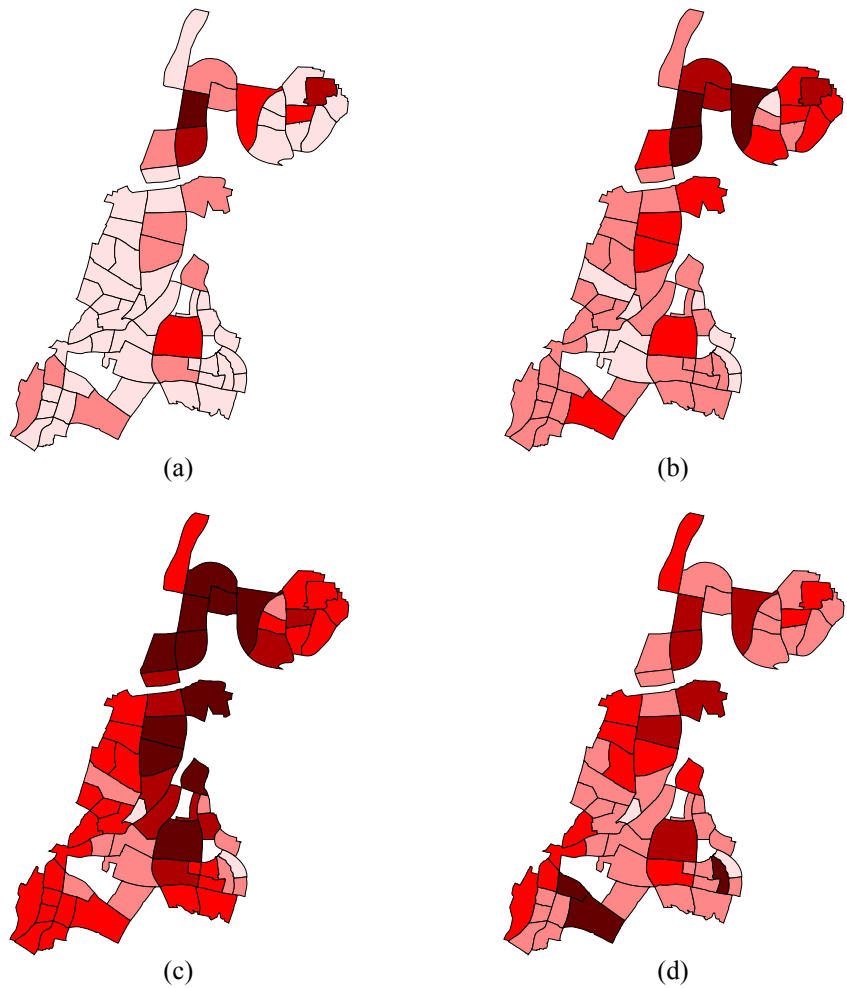

Figure 7: Betweenness (a), degree (b), closeness (c) and eccentricity (d) of the parks
(Note: values increase from light red to dark red)

The case study has demonstrated the modeling capability in the kind of structural analysis from both local and global perspectives. Throughout the paper, we have tried to show: (1) how the simplicial complex is complementary to a graph representation, both geometrically and algebraically, and (2) how the centrality measures, initially developed in the field of social networks or graph theory in general, can significantly contribute to the characterization of the status of individual simplices of a complex.

**6. CONCLUSION**
This paper introduced a model based on Q-analysis for exploring spatial relationships through common contextual objects. Through overlapping two map layers (a primary layer and a contextual layer), a simplicial complex can be set up for exploring various structural properties both from local and global perspectives. The major contribution of this paper is two-fold. First, the concept of spatial topology is suggested to extend the existing topological concepts and to concentrate on the network view of a geographic system for more advanced and robust spatial analysis in terms of information flow or substance exchange. Second, the three centrality-based measures are introduced for characterizing structural properties of individual objects within a geographic system. Thus it extends the conventional Q-analysis from initial qualitative analysis to quantitative analysis and enriches the capacity of Q-analysis to describe the topological status of individual objects in a given geographic system. Unlike their counterparts using a graph representation, the measures suggested here consider multidimensional chains of connectivity based on the concept of simplicial complex. Thus the model is, to a great extent, complementary to existing graph approaches in modeling information flow within a geographic system, or any natural and artificial system which can be abstracted as a network.

The model introduced is compatible to existing GIS model and thus is directly computable using layered geographic information commonly available in GIS. It is important to note that the analysis based on geometric representation is mainly applicable to a small system that is represented as a low dimensional complex, while the analysis using an algebraic approach, in particular the newly introduced measures, can be used for a relative large system that is treated as a higher dimensional complex. We plan to implement the model into ArcGIS environment in the near future, which will significantly enhance GIS functionality in topological analysis. We have seen the modelling advantage of our model in comparison to a graph approach. However, whether or not the measures could be good indicators for real world flow prediction is out of the scope of this paper. It certainly deserves further study in the future. From a broader perspective, we also plan to explore the relevance of this approach for a variety of geographical systems.

**ACKNOWLEDGEMENT**
Earlier versions of this paper were presented at the *9th AGILE Conference on Geographic Information Science*, 20 - 22 April, 2006, Visegrád, Hungary, and the *9th International Conference on Urban Planning and Urban Management*, 29 June - 1 July 2005, UCL, London.